\documentclass
[aps,prb,twocolumn,floatfix,english,showpacs,10pt,superscriptaddress]{revtex4-2}%
\usepackage{graphicx}
\usepackage{amsmath}
\usepackage{physics}
\usepackage{amssymb}
\usepackage{colordvi}
\usepackage{verbatim}
\usepackage{xcolor}
\usepackage{mathrsfs}
\usepackage{epsfig}
\usepackage{lipsum}
\usepackage{amsfonts}
\usepackage[unicode=true, breaklinks=false, pdfborder={0 0 1}, backref=false,
colorlinks=true, linkcolor=blue, urlcolor=blue, citecolor=blue]{hyperref}%
\providecommand{\U}[1]{\protect\rule{.1in}{.1in}}
\setcitestyle{numbers,square}
\begin{document}

\title{Soft Magnons in Anisotropic Ferromagnets}
\author{G.E.W. Bauer}

\affiliation{WPI Advanced Institute for Materials Research, Tohoku University, 2-1-1, Katahira, Sendai 980-8577, Japan}
\affiliation{Institute for Materials Research and CSIS, Tohoku University, 2-1-1 Katahira, Sendai 980-8577, Japan}
\affiliation{Kavli Institute for Theoretical Sciences, University of the Chinese Academy of Sciences, Beijing 10090, China}

\author{P. Tang}

\affiliation{WPI Advanced Institute for Materials Research, Tohoku University, 2-1-1, Katahira, Sendai 980-8577, Japan}

\author{M. Elyasi}

\affiliation{WPI Advanced Institute for Materials Research, Tohoku University, 2-1-1, Katahira, Sendai 980-8577, Japan}

\author{Y.M. Blanter}

\affiliation{Kavli Institute of NanoScience, Delft University of Technology, 2628 CJ Delft, the Netherlands}

\author{B.J. van Wees} 

\affiliation{Physics of Nanodevices, Zernike Institute for Advanced Materials, University of Groningen, 9747 AG Groningen, the Netherlands}

\date{\today }

\begin{abstract}
We discuss spin-wave transport in anisotropic ferromagnets with an emphasis on the zeroes of the  band edges as a function of a magnetic field. An associated divergence of the magnon spin should be observable by enhanced magnon conductivities in non-local experiments, especially in two-dimensional ferromagnets.

\end{abstract}
\maketitle

\section{Introduction}

\textquotedblleft Magnonics\textquotedblright\ is the study of the elementary
excitations of the magnetic order, i.e. spin waves and their quanta called
\textquotedblleft magnons\textquotedblright \cite{MagnonicsTB1,MagnonicsTB2,MagnonicsTB3}. It is believed to be competitive in future
information, communication, and thermal management technologies
\cite{Magnonics,SWC}. 

In the exchange interaction-only continuum model the spin wave dispersion is a parabola that shifts linearly with an applied magnetic field \cite{MagnonicsTB1,MagnonicsTB2,MagnonicsTB3}. The gain of
angular momentum of the ground state by flipping a single electron is
$\hbar$. The associated change of the magnetic momentum is $-2\mu_{B}$, where
$\mu_{B}$ is the Bohr magneton. The exchange energy cost of a single spin flip
is minimized by spreading this excitation over the whole system, forming a spin wave or its quantum, the magnon. 

Magnetic dipolar interactions and spin-orbit interactions strongly affect the spin wave dispersion of ferromagnets. Crystal anisotropies are the main
consequence of the latter, and cause, for example, magnon gaps in the absence
of an applied magnetic field. Only quite recently have researchers realized that
the magnon spin is not a universal constant. Ando \textit{et al}. \cite{Ando} reported enhanced spin
pumping by an elliptic magnetization precession, which, as we show below, can be interpreted as an enhanced magnon spin. Flebus \textit{et al}. \cite{Flebus} reported that the exchange magnon
polaron, i.e. the hybrid state of a magnon and a phonon, carries a spin
between $0$ and $\hbar$. Kamra and Belzig \cite{Kamra} predict
super-Poissonian shot noise in the spin pumping from ferromagnets into metallic
contacts based on a magnon spin that is enhanced from its standard value of
$\hbar$ by anisotropy \textquotedblleft squeezing\textquotedblright. These
authors start from a lattice quantum spin Hamiltonian with\ local and
magneto-dipolar anisotropies and predicted magnon spins of around $4 \hbar$ for the fundamental (Kittel) mode of an iron film. Kamra et al.  \cite{Kamra2} address the emergence of the magnon spins in antiferromagnets at weak applied magnetic fields. Neumann \textit{et al}.
\cite{Neumann2020} introduced an \textquotedblleft orbital contribution\textquotedblright    
to the magnon magnetic moment. Yuan et al. \cite{Yuan2022} review the history of the magnon spin concept and its enhancement by quantum squeezing 

Magnon currents can be injected into ferromagnetic insulators by heavy metal contacts, electrically by means of the spin Hall effect or by thermal gradients (spin Seebeck effect). \textit{Vice versa}, magnons can pump a spin current into a heavy metal contact and be detected by an inverse-spin-Hall
voltage. Both effects may be combined to study magnon transport in magnetic
insulators \cite{Cornelissen2015}. Films of ferrimagnetic yttrium iron garnet
are well suited for magnon transport studies and can be grown with high
quality down to a few monolayers \cite{Wei}. The same technique
also works well for antiferromagnetic hematite \cite{Lebrun,Wimmer,Liu}.

 de Wal \textit{et al}. \cite{Dewal} studied non-local magnon
transport in an antiferromagnetic van der Waals film with a perpendicular Néel vector. An in-plane magnetic field cants the
two sublattices until the material becomes ferrimagnetically ordered at the \textquotedblleft spin-flip\textquotedblright\ transition, not unlike the in-plane spin texture of hematite at high fields. In the absence of additional anisotropies, the band gap of the spin wave dispersion vanishes at this point,
i.e. the magnons become \textquotedblleft soft\textquotedblright.

\begin{figure}[t]
\centering
\includegraphics[width=1.0\linewidth]{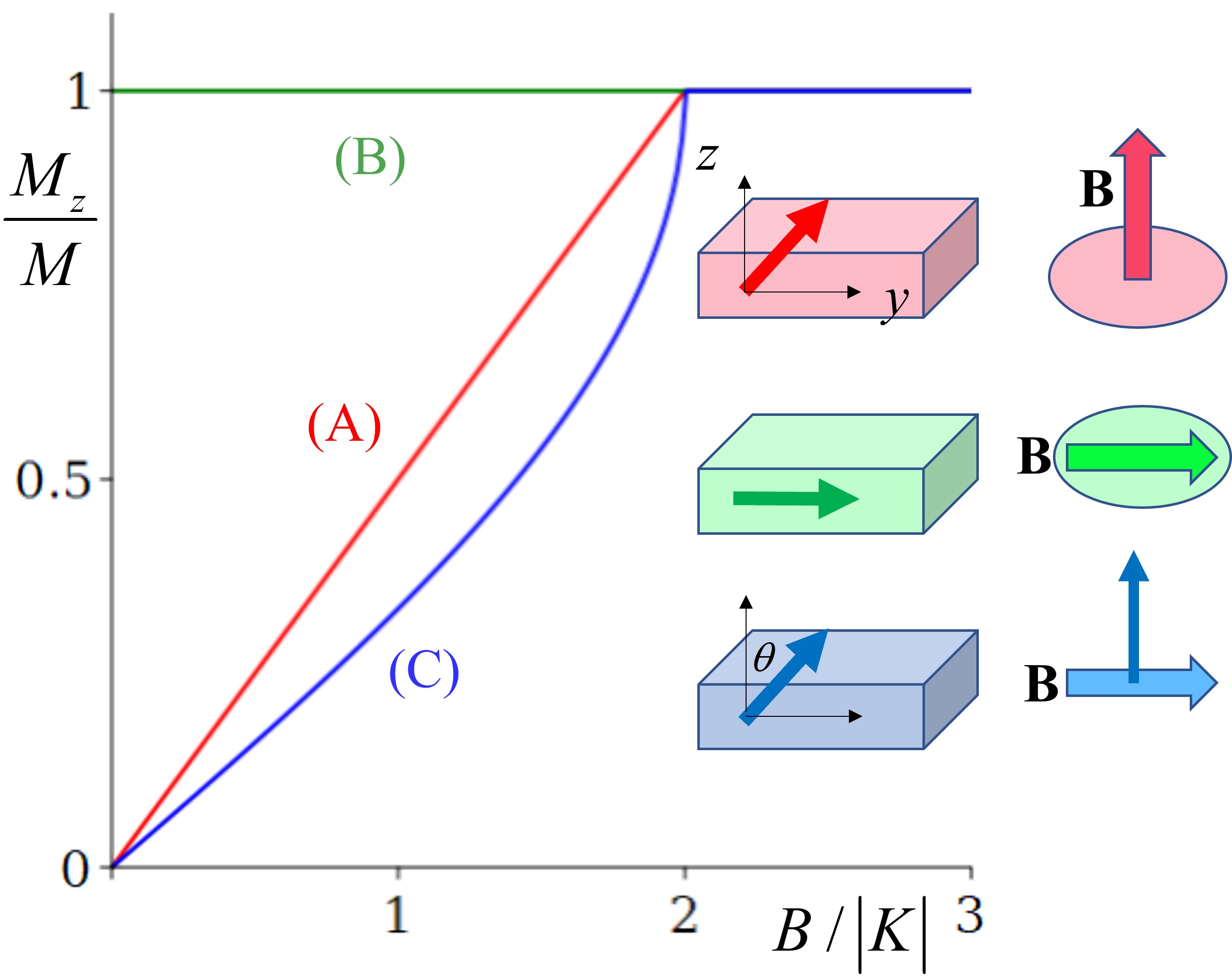} 
\caption{Three configurations of ferromagnets with uniaxial magnetic anisotropies parameterized by the constant $K$ and their magnetization $M_B$ in the direction of the applied magnetic field $B$. Case (A) is an easy-plane ferromagnet with a magnetic field along $z$ (red). In case (B) the field lies in the easy plane (green). In case (C) the field is normal to
the easy \textit{z}-axis and $\theta$
is the tilt angle of the magnetization. }%
\end{figure}

Here we make a step back by realizing that magnons can be \textquotedblleft
soft\textquotedblright\ in simple ferromagnets as well. We find that the magnon spin
can be strongly affected by the associated non-analyticities of the dispersion relation and may even diverge. We illustrate the general concept of \textquotedblleft soft magnons\textquotedblright\ and \textquotedblleft magnon
spin\textquotedblright\ for the three generic magnetic configurations
in Figure 1 in which applied magnetic fields and uniaxial anisotropies compete. We predict observable enhancements of magnon transport in the proximity of the soft magnon configurations. These results may partly explain the strong enhancement of the magnon transport at the spin-flip transition in anti-ferromagnets \cite{Ping}.

Sections II and III set the stage by re-deriving results for the ground and
excited states of long energy excitations in magnetic systems. In Section II we
analyze the ground state by minimizing the classical magnetic free energy. We
solve the linearized Landau-Lifshitz equations in Section 3 for the spin wave
dispersion relations, finding results that agree with those obtained from
quantum spin models. In Section IV we address the magnon spin in a way we have
not found in the literature. We show that the Hellman-Feynman theorem can be a
useful tool to get hands on the magnon magnetic moment, but only when field
and magnetization are collinear. In Section V we discuss possible experimental
signatures of the large magnon spins close to the kinks in the magnon
dispersion. Section VI contains a critical discussion of the model and recommendations for future work.

\section{Spin Hamiltonians and ferromagnetic ground states}

We consider Hamiltonians for local spins $\hat{\mathbf{S}}_{i}$ and magnetic
moments $\mathbf{\hat{M}}_{i}=-\gamma\hat{\mathbf{S}}_{i}$ on lattice sites
$i$:
\begin{equation}
\mathcal{H}=\gamma\hbar\sum_{i}\hat{\mathbf{S}}_{i}\cdot\mathbf{B}-\sum
_{ij}J_{ij}(\hat{\mathbf{S}}_{i}\cdot\mathbf{\hat{S}}_{j})-\bar{K}\sum
_{i}(\hat{S}_{i}^{z})^{2}, \label{Ham}%
\end{equation}
where $J_{ij}$ is the exchange integral between spins on sites $i \ne j$,
$-\gamma$ is the gyromagnetic ratio of an electron, and $\mathbf{B}$ is a constant external magnetic field. We focus here
on $J_{ij}>0$ that leads to ferromagnetic order. $\bar{K}$ is a uniaxial
anisotropy parameter along the Cartesian $z$-direction, $\bar{K}>0$ $\left(
\bar{K}<0\right)  $ corresponds to an easy-axis (easy-plane) ferromagnet.

The ground state that minimizes the total energy of the macroscopic system is ferromagnetic. The total spin of the system $\hat{\mathbf{S}%
}=\sum_{i}\hat{\mathbf{S}}_{i}$  then becomes a classical vector \(\mathbf{S}\) corresponding to a magnetization
density $\mathbf{M}=-(\gamma\hbar/\Omega)\mathbf{S}$, where $\Omega$ is the crystal volume. The associated energy density without irrelevant constant terms reads
\begin{equation}
E(\mathbf{M})=-\mathbf{M}\cdot\mathbf{B}-\frac{\bar{K}}{M}(M_{z})^{2}+\frac{D}
{2M}\left(  \boldsymbol{\nabla}\mathbf{M}\right)  ^{2} + E_{\mathrm{dmag}}(\mathbf{M}), \label{Energ}%
\end{equation}
where $M=\left\vert \mathbf{M}\right\vert =(\gamma\hbar S/\Omega)$ and $ S=\left\vert \mathbf{S}\right\vert$. The spin-wave stiffness $D$  represents the exchange energy cost of spatial deformations in the continuum limit that depends on crystal structure and the exchange parameters \(J_{ij}\).  The magnetizations of the
ground states are constant in space with zero exchange energy. The demagnetization energy in magnetic films with surface normal \(\mathbf{n}\) along \(z\) can then be absorbed into the anisotropy field
$K=(S\bar{K})/(\gamma\hbar) + M$. We discuss here three configurations: (A) Easy $xy$-plane anisotropy $(K<0)$ and
the magnetic field normal to the plane, (B) easy $xy$-plane anisotropy with an
in-plane magnetic field (Kittel problem), and (C) easy $z$-axis anisotropy $(K>0)$
with a magnetic field in the \textit{y}-direction. Other configurations such as in-plane easy axis etc. give similar results. 

We are interested in
the discontinuities that emerge at critical fields $B_{c}=0$ for case (B) and
$B_{c}=2\left\vert K\right\vert $ for (A) and (B). 

In case (A) $\mathbf{B}=B\mathbf{z},$ where $\mathbf{z}$ is the unit vector
along the anisotropy axis, the energy and magnetizations are discontinuous at $B_{c}=2\left\vert K\right\vert $
\begin{equation}
\frac{E_{0}^{\left(  A\right)  }}{M}=\left\{
\begin{array}
[c]{c}%
-\frac{B^{2}}{4\left\vert K\right\vert }\\
\left\vert K\right\vert -B
\end{array}
\text{    for    }%
\begin{array}
[c]{c}%
B<2\left\vert K\right\vert \\
B>2\left\vert K\right\vert
\end{array}
\right.
\end{equation}
\begin{equation}
\frac{M_{z}}{M}=\left\{
\begin{array}
[c]{c}%
\frac{B}{2\left\vert K\right\vert }\\
1
\end{array}
\text{   for   }%
\begin{array}
[c]{c}%
B<2\left\vert K\right\vert \\
B>2\left\vert K\right\vert
\end{array}
\right.
\end{equation}

In case (B) the energy
\begin{equation}
E^{\left(  B\right)  }(\mathbf{M})=-M_{y}B+\frac{\left\vert K\right\vert }
{M}M_{z}^{2}%
\end{equation}
is minimal for $\mathbf{M}_{0}=\left(  0,M,0\right)  $ for all $B\neq0.$
Finally, in (C) a field $\mathbf{B}=B\mathbf{y\ }$along the $y$-axis tilts the
magnetization into the $yz$-plane with $\mathbf{M}_{0}=M(0,\sin\theta
,\cos\theta)$, where $\theta$ is the angle with the out-of-plane direction.
Eq. (\ref{Energ})
\begin{equation}
\frac{E^{\left(  C\right)  }(\theta)}{M}=-K\cos^{2}\theta-B\sin\theta.
\end{equation}
is minimized for%
\begin{equation}
\sin\theta=\left\{
\begin{array}
[c]{c}%
\frac{B}{2K}\\
1
\end{array}
\right.  \text{ for }%
\begin{array}
[c]{c}%
B<B_{c}\\
B>B_{c}%
\end{array}
.
\end{equation}
Above $B_{c}=2K,$ the field and magnetization are aligned.

\section{Spin wave dispersion}

Here we consider the frequency dispersion relation for the elementary excitations in homogeneous extended magnets that can be bulk crystals, thin films, or two-dimensional systems. The excitation frequencies are sharply defined in the limit of small amplitude oscillations, in which spin waves can be mapped on a set of non-interacting harmonic oscillators. The magnon Hamiltonian is the lowest-order term in the Holstein-Primakoff expansion of the spin Hamiltonian, which can subsequently be diagonalized by a Bogoliubov transformation \cite{MagnonicsTB1,MagnonicsTB2,MagnonicsTB3}.

Here we chose to start from the Landau-Lifshitz equation $\mathbf{\dot{M}%
}=-\gamma\left(  \mathbf{M}\times\mathbf{B}_{\mathrm{eff}}\right)  $, in which
$\mathbf{B}_{\mathrm{eff}}=-\partial E\left(  \mathbf{M}\right)
/\partial\mathbf{M}$ and $E\left(  \mathbf{M}\right)  $ is the classical
magnetic free energy introduced above. Writing \(\mathbf{M}_q=\mathbf{M}_{0}+\mathbf{m}_{q}\) and to leading order in the small transverse excitation amplitudes $\mathbf{m}_{q}
\cdot\mathbf{M}_{0}=0$,  the spin wave frequencies $\omega_q$ for wave vector \(\mathbf{q}\) are the solutions  of
\begin{equation}
\left(  i\omega_q-\mathbf{B}_{\mathrm{eff}}\times\right)    \mathbf{M}_{q} =\mathbf{0}.
\end{equation}
For simplicity, we compute \(\omega_q\) without dipolar corrections that cause well-known anisotropic corrections for 
\(q \ne 0\), but do not affect the band edges in Figure 3. The results agree with those obtained from the lowest-order Holstein-Primakov expansion of the corresponding spin Hamiltonians.

\begin{figure}[t]
\centering
\includegraphics[width=1.0\linewidth]{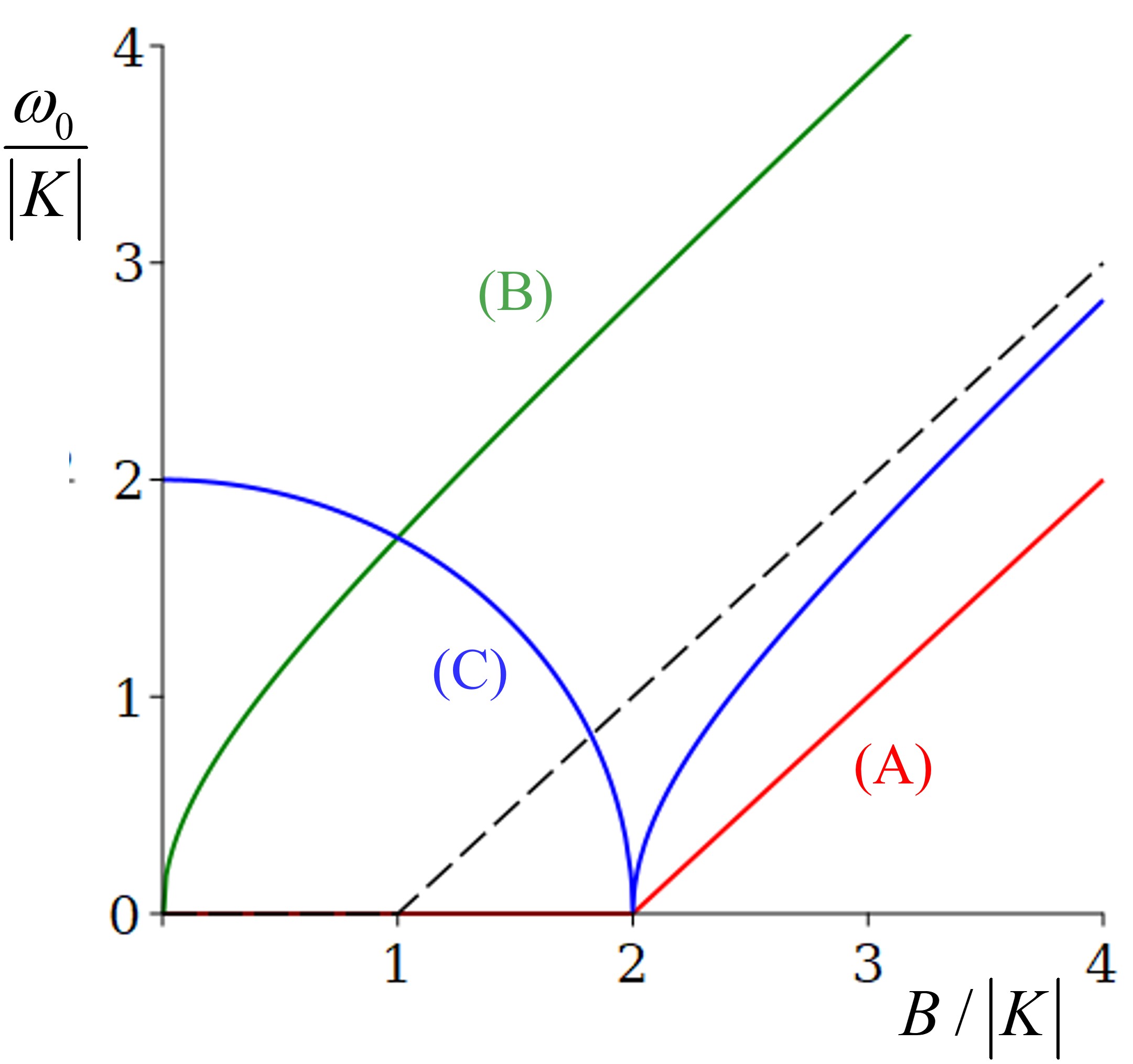} 
\caption{Spin wave frequency of the magnon band edges for the configurations (A)-(C) in Fig. 1. The dashed black line \(\omega_0 / \gamma = B -K \)  is the asymptotic form of the case (B) at large fields.}
\label{Figure2}%
\end{figure}

When the magnetic field is normal to the easy plane (A), the spin wave dispersion reads
\begin{equation}
\omega_{q}^{\left(  A\right)  }=\gamma\left\{
\begin{array}
[c]{c}%
\sqrt{Dq^{2} \left( 2\left\vert K\right\vert - \frac{B^2}{4K^2}+Dq^{2}\right)}\\
B-2\left\vert K\right\vert+ Dq^{2}%
\end{array}
\text{  for  }%
\begin{array}
[c]{c}%
B<2\left\vert K\right\vert \\
B>2\left\vert K\right\vert
\end{array}
\right.
\end{equation}

The spin wave frequency of the uniform precession (\(q=0)\) vanishes for magnetic fields below $B_{c}$ because the torques generated by the anisotropy and applied fields cancel. The energies for all magnetization directions that lie on a cone with opening angle \(\theta\) are the same, so the ground state is highly degenerate.

The Kittel problem (B) leads to 
\begin{equation}
\omega_{q}^{\left(  B\right)  }=\gamma\sqrt{\left(  B+Dq^{2}+2K\right)
\left(  B+Dq^{2}\right)  }.
\end{equation}
The anisotropy qualitatively changes the dispersion at small magnetic fields
from $\omega_{0}^{(B)}=\gamma B$ for $K=0$ to $\omega_{0}^{(B)}\sim\gamma\sqrt{B}$ when
$K>B$. The anisotropy breaks the axial symmetry and mixes the anti-clockwise circular precession mode with positive frequencies and the forbidden clockwise one with negative frequencies, which leads to the square root dependence rather than $2\gamma K$, the linearly extrapolated value from the high-field region. 

In configuration (C) (as in (A)),  magnetization and field are not
collinear for fields $B<B_{c}$. By rotating the coordinate system around the $x$-axis by $\theta$
such that $\mathbf{M}_{0}^{\prime}$ is along $z'$, we can
impose the magnon approximation by solving for small amplitude oscillations
$\mathbf{m}_{q}^{\prime} \cdot \mathbf{M}_{0}^{\prime}=0$. The result is
\begin{align}
\omega_{q}^{\left(  C\right)  }  &  =\gamma\left\{
\begin{array}
[c]{c}%
\sqrt{\frac{1}{2K}\left(  2K+Dq^{2}\right)  \left(  4K^{2}-B^{2}+2DKq^{2}\right)} \\
\sqrt{\left(  B+Dq^{2}-2K\right)  \left(  B+Dq^{2}\right)  }%
\end{array}
\right. \nonumber\\
&  \text{ for }%
\begin{array}
[c]{c}%
B<2\left\vert K\right\vert \\
B>2\left\vert K\right\vert
\end{array}
.
\end{align}
We observe that the \(B\)-dependence of the collinear configuration equals that of the Kittel mode shifted by $2\left\vert K\right\vert$ to higher magnetic fields.

\section{Magnon spin and Hellmann-Feynman theorem}

An excited state $\left\vert q\right\rangle $ with energy $\varepsilon_{q}=\hbar \omega_q$ of an arbitrary
spin Hamiltonian carries a spin magnetic moment $\boldsymbol{\mu}_{q}=\left\langle
q\right\vert \mathbf{\hat{M}}\left\vert q\right\rangle -\mathbf{M}_0,$ where $\hat {\mathbf{M}}=-\gamma\hbar\sum_{i}\hat{\mathbf{S}}_{i}.$ Consider a spin system with zero-field Hamiltonian $\mathcal{H}_{S}$ that interacts with constant applied magnetic field $\mathbf{B}$ by the Zeeman interaction
\begin{equation}
\mathcal{H}=\mathcal{H}_{S}-\mathbf{\hat{M}}\cdot\mathbf{B} .
\end{equation}
The Hellmann-Feynman theorem then states that for simplicity in the absence of textures,
\begin{equation}
\boldsymbol{\mu}_{q}=-\frac{\partial\varepsilon_{q}}{\partial B_{\Vert}}%
\frac{\mathbf{M_0}}{M}  \label{HFcoll}
\end{equation}
where $B_{\Vert}=\mathbf{B}\cdot \mathbf{M}_0/M$. For our classical spin system, we  replace $\varepsilon_{q}$ with $\Delta E_q =E(\mathbf{M}_0+\boldsymbol{\mu}_q)-E(\mathbf{M}_0)$.
When the magnetization and the applied field are parallel, we can simply read off the spin of the excited state from the magnon spectrum as a function of the applied field. When $\mathbf{M}\nparallel\mathbf{B}$, the situation is more complicated and the Hellmann-Feynman theorem requires additional calculations. Configurations (A) and (C) at fields $B<2K,$ for example, acquire an
\textquotedblleft orbital correction\textquotedblright\ \cite{Neumann2020}\
\begin{equation}
\mu_{q}=-\frac{d\varepsilon_{q}}{dB}+\frac{\partial\varepsilon_{q}}
{\partial\theta}\frac{\partial\theta}{\partial B},  \label{HFnoncoll}
\end{equation}
where $\theta$ is the equilibrium tilt angle. The implicit dependence on the field enters here with an opposite sign compared to Ref. \cite{Neumann2020}. 

 To leading order, the solutions of the LL equation are transverse, i.e. $\mathbf{m}_{q}\cdot\mathbf{M}_{0}=O(m_q^2)$.  If we transform back into the time domain, the transverse components oscillate with the spin wave frequency and average out to zero. A magnon moment along the magnetization direction persists in the time-average as
\begin{equation}
\boldsymbol{\mu}_{q}\approx-\frac{1}{2}\left(  \frac{\left\vert
\mathbf{m}_{q}\right\vert }{M}\right)  ^{2} \frac{\mathbf{M}_{0}}{M} .
\end{equation}

The amplitudes of the solutions of the LL equation are continuous but can be
quantized by requiring that the minimum excitation energy of a mode with
energy $\varepsilon_{q}$ is discrete:
\begin{align}
\Delta E_q & = E (\mathbf{M}_q)-E_0 = \hbar \omega_{q}.
\end{align}
By normalizing the mode amplitudes in this way we can compute the spin of a
single magnon for our three configurations (A-C).

\begin{figure}[t]
\centering
\includegraphics[width=1.0\linewidth]{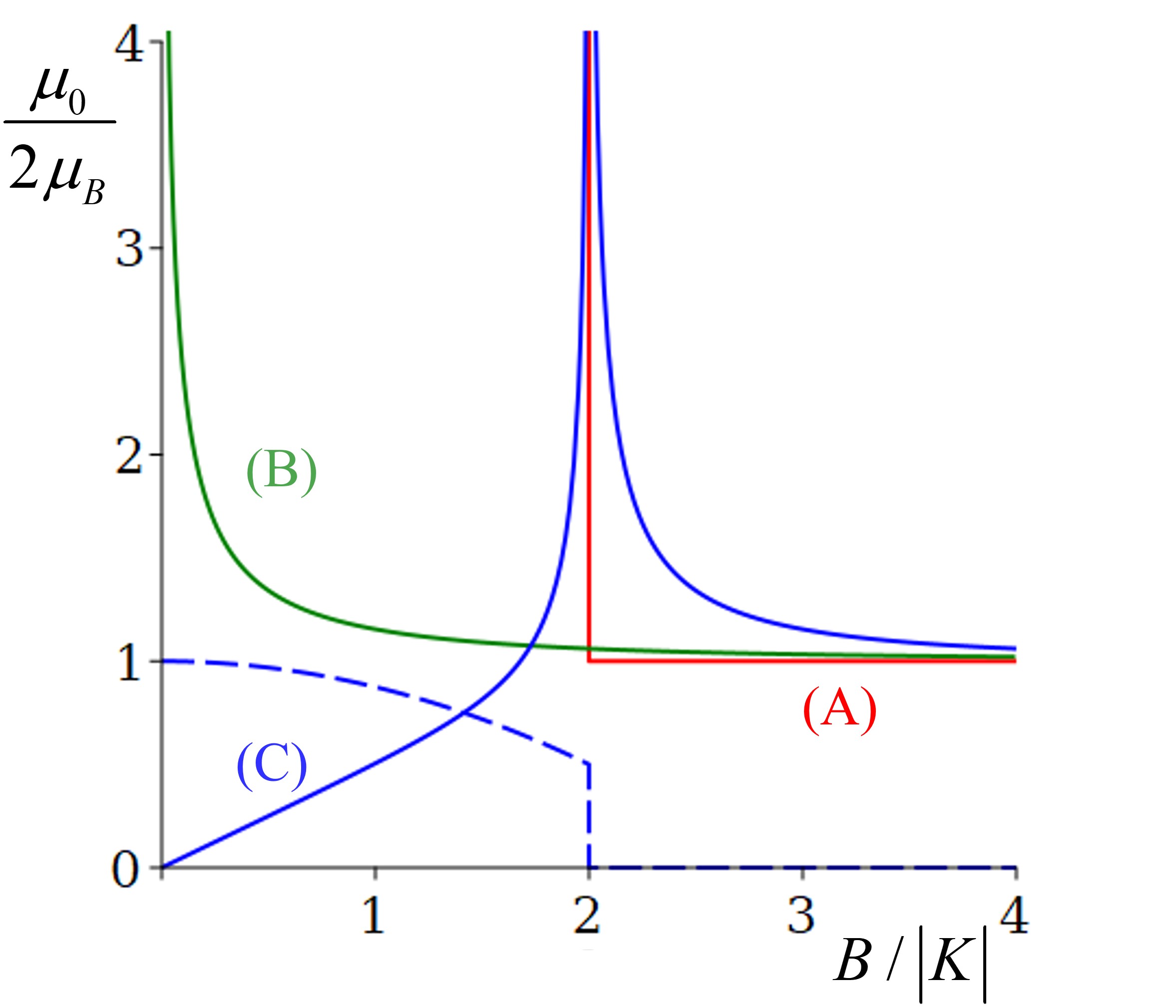} 
\caption{Magnon spin at
the band edges for the configurations (A)-(C) in Fig. 1. In (C) the magnon
spin is tilted for $B<2\left\vert K\right\vert $ with a component in the $y$-
(full curve) and $z$-directions (dashed curve).}%
\label{Figure3}%
\end{figure}

In case (A), the moment of a single magnon $\mu_{q}^{(A)}=- 2\mu_{B}$ for $B>B_{c}$ from the Hellmann-Feynman. However, spin and magnetization are not collinear for $B<B_{c}$. The LL equation then leads to a magnon magnetic moment along the equilibrium magnetization direction of
\begin{equation}
\frac{\mu_{q}^{\left( A\right)  }}{2\mu_{B}}= - \frac{|K|-\frac{B^2}{4|K|}+Dq^{2}}{\sqrt{Dq^2(2|K|-\frac{B^2}{2|K|}+Dq^{2})}}.
\end{equation}
that diverges when \( q \rightarrow 0\) because the degeneracy of all magnetization directions on the cone with angle \(\theta\) allows coherent precession of the magnetization without an energy cost. 

In case (B), we can compute the magnon spin simply as a derivative of the frequency with respect to the applied magnetic field from the Hellmann-Feynman theorem Eq. (\ref{HFcoll}) since \(\mathbf{B} \parallel \mathbf{M}\) : 
\begin{equation}
\frac{\mu_{q}^{\left( B\right)  }}{2\mu_{B}}= - \frac{B+K+Dq^{2}}{\sqrt{\left(
B+Dq^{2}+2K\right)  \left(  B+Dq^{2}\right)  }}.
\end{equation}
In the zero field limit $\mu_{0}^{\left(  B\right)  }=-2\mu_{B}/\sqrt{K/(2B)}$
diverges.  The anisotropy mixes right and left precession modes to create an elliptic motion that in the limit of zero magnetic field leads to a linear \(x\)-polarized motion in the easy plane. As in case (A) the restoring torque vanishes in this limit, which allows the magnon amplitude to become large.

Also in the high field regime $B>B_{c}$ of the case (C), we can use the Hellmann-Feynman theorem without having to worry about \textquotedblleft orbital\textquotedblright\ terms 
\begin{equation}
\frac{\mu_{q}^{\left(  C\right)  }}{2\mu_{B}}=-\frac{\left(  B+Dq^{2}\right)
+\left(  B+Dq^{2}-2K\right)  }{\sqrt{\left(  B+Dq^{2}-2K\right)  \left(
B+Dq^{2}\right)  }}\text{ for }B>B_{c},
\end{equation}
which is a shifted version of $\mu_{q}^{\left(  B\right)  }$.  $\mu_{0}^{\left(  C\right)  }$ diverges for $B \downarrow B_c$ because the torque in the out-of-plane \(z\)-direction vanishes.  The magnon spin for the canted configuration $\left(  B<B_{c}\right)  $ can 
be computed directly from the solutions of the LL equation. Along the canted
equilibrium magnetization direction
\begin{gather}
\frac{\mu_{q}^{\left(  C\right)  }}{2\mu_{B}}=-\frac{\sqrt{Dq^{2}+2K}}
{\sqrt{2K}}\times\nonumber\\
\frac{\sqrt{4K^{2}-B^{2}+2DKq^{2}}\left(  8K^{2}-B^{2}+4DKq^{2}\right)
}{(4K^2-H^2)4K+(16K^2+4KDq^2-3H^2)Dq^2}
\end{gather}
which at the band edge \(q=0\) diverges like \(1/\sqrt{B_c-B}\). In Figure 3 we plot the in-plane projection of the magnon spin at the band edges along the $y$-direction, i.e. the spin component $\mu_{0}^{\left(
C\right)  }\sin\theta$ that can be injected and detected by Pt contacts, as well as the $z$-component along the easy axis. 

\section{Transport}

We now discuss some physical consequences of the results derived above.

In case (A) the magnons are gapless up to $B_{c}=2\left\vert K\right\vert .$
The absence of an energy cost of the in-plane magnetization amplitude implies that the equilibrium direction is arbitrary and depends on the history, remaining in-plane anisotropies, or disorder.  
Even small energy gains of the magneto-dipolar interactions break a uniform magnetization down into domains. At $B_{c}$ the magnon spin jumps to its standard value of \(-2\mu_B\). In case (B) interesting effects may occur when the magnon spin diverges at vanishing magnetic fields but it is again difficult to control the equilibrium magnetic order. Therefore, the soft magnon in
the field dependence of case (C) at the critical field looks most interesting. While the fluctuations
and the associated magnon spin become large, the finite applied field hinders
the breaking of the magnetic order even when the magnon gap vanishes. Hence, we focus on case (C) with an applied magnetic field that
approaches the critical value from above. We address magnon transport in bulk
materials and in two-dimensional ferromagnets.

\subsection{Spin pumping}

The magnon spin affects the spin pumping \cite{Ando} and spin pumping shot
noise \cite{Kamra}. The spin pumped into the metal by the dynamics of an
insulating or metallic ferromagnet in units of ampere reads
\begin{equation}
\mathbf{J}_{s}=\frac{e}{2\pi}\left[  \frac{g_{r}}{M_{0}^{2}}\mathbf{M}%
\times\mathbf{\dot{M}+}\frac{g_{i}}{M_{0}}\mathbf{\dot{M}}\right]
\end{equation}
where $g_{r}+ig_{i}$ is the dimensionless interface spin-mixing conductance
\cite{Tserkovnyak}. Inserting the LL equation for case (C) and $B\downarrow B_{c}$ leads to a dc spin current:
\begin{align}
\mathbf{J}_{s}^{\left(  \mathrm{dc}\right)  } & =-\frac{e}{2\pi}\gamma\frac
{g_{r}}{M_{0}^{2}}\left(  \mathbf{M}_{0}\cdot\mathbf{B}_{\mathrm{eff}}\right)
\boldsymbol{\mu}_{0} \nonumber \\
& =\frac{eg_{r}}{2\pi}\frac{N}{M_{0}}\mu_{0}\omega_{0}%
=\frac{eg_{r}}{2\pi}\frac{2\mu_{B}N}{M_{0}}\gamma B_{c}\mathbf{.}%
\end{align}
where $N$ is the number of magnons in the Kittel mode. Since the divergence
of the magnon spin is canceled by the vanishing resonance frequency,
we do not expect any anomalies of the spin pumping when the magnons become soft.

\subsection{Magnon conductivity and spin Seebeck effect}

Next, we consider diffuse dc magnon transport in a bulk ferromagnet under gradients of temperature or magnon chemical potential with a special focus on  the magnon conductivity of the magnon soft mode in configuration (C).
We employ the relaxation time approximation to the linearized Boltzmann equation in which the magnon current is polarized along the equilibrium magnetization
$\mathbf{M}_{0}.$ We disregard the magneto-dipolar interactions on the dispersion relation since these
cancel to a large extent in magnon transport except at very low
temperatures.  The magnon conductivity $\sigma_{m}^{\left(  nd\right)  }$ in electric units of Sm\(^{2-n}\) \cite{Wei} and spin
Seebeck coefficient $S_{m}^{\left(  nd\right)  }$ in units of  V/m then read
\begin{equation}
\sigma_{m}^{\left(  nd\right)  }=\frac{e^{2}\tau_{r}^{\left(  nd\right)  }%
}{k_{B}T}\int\frac{dq^{n}}{\left(  2\pi\right)  ^{n}}\frac{-\mu_{q}}{2\mu_{B}%
}\mathbf{v}_{q}^{2}\left(  f_{P}\right)  ^{2}e^{\frac{\hbar\omega_{\mathbf{q}%
}}{k_{B}T}}, \label{conductivity}%
\end{equation}
\begin{equation}
\sigma_{m}^{\left(  nd\right)  }S_{m}^{\left(  nd\right)  }=\frac{e\tau_{r}^{\left(  nd\right)  }}
{k_{B}T^{2}}\int\frac{dq^{n}}{\left(  2\pi\right)  ^{n}}\frac{-\mu_{q}}
{2\mu_{B}}\hbar\omega_{q}\mathbf{v}_{q}^{2}\left(  f_{P}\right)  ^{2}
e^{\frac{\hbar\omega_{\mathbf{q}}}{k_{B}T}},
\end{equation}
where $f_{P}=\left[  \exp\left(  \frac{\hbar\omega_{\mathbf{q}}}{kT}\right)
-1\right]  ^{-1}$ is the Planck distribution, $\mathbf{v}_{q}=\partial
\omega_{q}/\partial k_{y}$ is the group velocity in the transport
$y$-direction and $n$ is the spatial dimension.\ The average relaxation time
$\tau_{r}^{\left(  nd\right)  }$ can be very different in 2 and 3 dimensions even 
when everything else remains the same \cite{Wei}. Typical frequencies are in
the GHz regime, so assuming that temperatures at not too low, $k_{B}T\gg
\hbar\omega_{q}.$ The
combination $\mu_{q}\omega_{q}$ in the integrand of the spin Seebeck coefficient is analytic when the magnon turns soft. So we do not expect anomalies in the thermally driven spin transport.

An artifact of the above equations, derived for a continuum model of the lattice, a constant relaxation time approximation, and the high-temperature limit, is the divergence of Eq. (\ref{conductivity}) at high wave numbers. When $K=0$ and in three dimensions 
\begin{align}
\sigma_{m}^{\left(  3d\right)  } &=\frac{e^{2}}{\hbar^{2}}\frac{2}{3\pi^{2}}%
\tau_{r}^{\left(  3d\right)  }k_{B}T \nonumber \\ 
& \left(  Q_{\infty}+\frac{Bq}{2\left(
DQ_{\infty}^{2}+B\right)  }-\frac{3}{2}\sqrt{\frac{B}{D}}\arctan\!\left(
\sqrt{\frac{D}{B}}Q_{\infty}\right)  \right)  ,
\end{align}
where $Q_{\infty}$ is an ultraviolet cut-off, at room temperature provided by the onset of strong
magnon-phonon scattering at THz frequencies. Including the easy axis anisotropy $K>0$ 
causes a logarithmic divergence at the critical field $B\downarrow B_{c}$ that can be regulated by a low-momentum
cut-off $Q_{0}$ that can be rationalized by a residual in-plane anisotropy or disorder. To leading orders for large $Q_{\infty}$ and small $Q_{0}%
^{2}:$
\begin{equation}
\sigma_{m}^{\left(  3d\right)  }\rightarrow
\frac{e^{2}}{\hbar^{2}}\frac{2}{3\pi^{2}}\tau_{r}^{\left(
3d\right)  }k_{B}T
\left(  Q_{\infty}-\frac{\sqrt{2}}
{16}\sqrt{\frac{K}{D}}\ln\frac{DQ_{0}^{2}}{8K}\right)
\end{equation}
The last term is the conductivity enhancement by the soft mode magnon.

In two dimensions and $K=0$
\begin{align}
\sigma_{m}^{\left(  2d\right)  }&=\frac{e^{2}}{\hbar^{2}}\frac{1}{\pi}\tau
_{r}^{\left(  2d\right)  }k_{B}Te^{2}\tau_{r}^{\left(  2d\right)  }\frac
{k_{B}T}{\pi}\frac{1}{2} \nonumber \\ 
& \left[  \ln\left(  1+\frac{DQ_{\infty}^{2}}{B}\right)
+\frac{B}{B+DQ_{\infty}^{2}}\right]
\end{align}
the divergence is logarithmic, but with vanishing magnon gap $B\rightarrow0$ there is now also a logarithmic infrared divergence for a constant magnon spin, which is caused by the enhanced magnonic density of states at the band edge. Including the divergent magnon spin when \(K>0\), to
leading order in a small $Q_{0}$ and large $Q_{\infty}$, and for  $B\downarrow B_{c}$
\begin{equation}
\sigma_{m}^{\left(  2d\right)  }\rightarrow e^{2}\tau_{r}^{\left(  2d\right)
}\frac{k_{B}T}{\pi}\left[  \ln\left(  \sqrt{\frac{2D}{K}}Q_{\infty}\right)
+\sqrt{\frac{K}{2D}}\frac{1}{4Q_{0}}\right]
\end{equation}
The high-momentum divergence is still logarithmic, but the low-momentum one becomes
algebraic. This implies a dimensionally enhanced magnon transport around the critical magnetic field
in van der Waals ferromagnets with perpendicular magnetization.

\section{Discussion}

Several processes regulate the enhanced fluctuations that cause the divergence
of the magnon spin reported here, but should not destroy the predicted
enhancement of spin pumping and spin transport. The situation is not unlike
that of ferrimagnets with singular points in the phase diagram caused by
angular or magnetic momentum compensation that causes interesting enhancements
of e.g. domain wall velocities, in their vicinity \cite{Beach}.

The magnon spin of the Kittel mode can be observed by spin pumping under
ferromagnetic resonance. Ando \textit{et al.} \cite{Ando} indeed reported that
the ellipticity of the magnetization precession under ferromagnetic resonance
conditions increases the spin pumping, a finding that we interpret as evidence
for an enhanced magnon spin. However, the resonance frequency also enters the
pumped spin current, and the product spin$\times$frequency remains
well-behaved for soft magnons. Similarly, the spin Seebeck effect is hardly affected. 

Magnon transport using heavy metal contacts for spin injection and detection
can be carried out only on magnets that are also good electrical insulators. The
magnon conductivity can be measured reliably when the injector and detector
separation does not exceed the magnon diffusion length, which requires high
magnetic quality. YIG films can be tuned to a perpendicular magnetization by doping and epitaxial  strain  \cite{PPAYIG}. The hexagonal barium ferrite
(BaFe$_{12}$O$_{19}$) has a perpendicular anisotropy field of 1.7 teslas and damping
constant $\alpha=10^{-3}$ \cite{BMO}.  Other options are electrically insulating van der Waals ferromagnets with perpendicular magnetization such as CrI\(_3\) \cite{Huang}. With minor adaptions, the  present formalism can be used as well for magnetic films (strips) with an in-plane easy crystal (demagnetization) anisotropy axis and an in-plane magnetic field applied at right angles to it.

We estimate the magnitude of the expected effects by adopting the spin wave stiffness of YIG $D=5\cdot10^{-17}\,\mathrm{Tm}^{2},$ a perpendicular anisotropy
$K=1\,\mathrm{T,}$ a high momentum cut-off frequency of $1\,\mathrm{THz}$ and
an in-plane anisotropy of $1\,\mathrm{mT.}$ This leads to an enhancement of
the magnon conductivity at the soft magnon point of  \(\sim 2\) for
three-dimensional and \(\sim 5\) for two-dimensional magnets. 

The divergences reported here occur only in materials with
weak residual anisotropies, \textit{i.e.} a sufficiently small cut-off \(Q_0\). Moreover,  non-parabolicities render the magnon approximation invalid when the spin wave amplitudes become large by large magnon spins and/or numbers. The implied magnon interactions may act as a brake on
the\ dynamics. The predicted numbers at the critical points should therefore be taken with a grain of salt, but the enhancement of the magnon thermal conductivity close to the softening of the magnon modes should persist even when these factors are taken into account.

\section{Conclusions}

We considered the spin waves of ferromagnetic magnons as
a function of an applied magnetic field, focusing on the singular points at
which the band edges (nearly) vanish. The more general conclusions such as the relation between the Hellmann-Feynman theorem for collinear systems also hold for antiferromagnets. We predict enhanced non-local magnon
transport caused by the divergence of the magnon spin for magnets with
a magnetic field applied perpendicular to a uniaxial magnetic anisotropy, which is stronger
in two than in three dimensions. The effect should contribute to the observation of magnon transport above and down to the \textquotedblleft spin-flip\textquotedblright\ transition at which the spin sublattices are forced to align ferromagnetically
  \cite{Dewal} and we expect enhanced non-local signals carried by the soft acoustic magnon as in case (C). However, more work is necessary to fully understand the experiments \cite{Ping}. 

The present calculations of the transport properties address the linear response at elevated temperatures, disregarding the effect of non-linear terms
that are likely to become important. Higher harmonic generation is easier for floppy magnetic order but is at vanishing applied fields complicated by spontaneous magnetic textures \cite{Koerner,Guoqiang}, which may not interfere when the soft magnon is shifted to sufficiently large magnetic fields.  It should also be interesting to study the non-linearities of propagating 
soft magnons by microwave spectroscopy
\cite{Haiming}.

\begin{acknowledgments}
P.T. and G.B. acknowledge the financial support by JSPS KAKENHI Grants No. 19H00645 and No. 22H04965. BW was supported by the 
European Union’s Horizon 2020 program under Grant Agreement No. 785219 and 881603,  the NWO Spinoza prize, and 
ERC Advanced Grant 2DMAGSPIN (Grant agreement No.
101053054). Akash Kamra kindly commented on the manuscript.
\end{acknowledgments}

\end{document}